\documentclass{sigchi-ext}
\usepackage[T1]{fontenc}
\usepackage{textcomp}
\usepackage[scaled=.92]{helvet} % for proper fonts
\usepackage{graphicx} % for EPS use the graphics package instead
\usepackage{balance}  % for useful for balancing the last columns
\usepackage{booktabs} % for pretty table rules
\usepackage{ccicons}  % for Creative Commons citation icons
\usepackage{ragged2e} % for tighter hyphenation
\usepackage{xcolor}
\usepackage{soul}

\def\plaintitle{Consent on the Fly: Developing Ethical Verbal Consent for Voice Assistants}
\def\plainauthor{William Seymour, Mark Cote, and Jose Such}

\def\plainkeywords{Voice assistants; consent; data protection; data sharing.}
\title{\plaintitle}

\numberofauthors{3}
\author{%
  \alignauthor{%
    \textbf{William Seymour}\\
    \affaddr{King's College London} \\
    \affaddr{London, UK} \\
    \email{william.1.seymour@kcl.ac.uk}} \vfil \alignauthor{%
    \textbf{Mark Coté}\\
    \affaddr{King's College London} \\
    \affaddr{London, UK} \\
    \email{mark.cote@kcl.ac.uk}} \vfil \alignauthor{%
    \textbf{Jose Such}\\
    \affaddr{King's College London} \\
    \affaddr{London, UK} \\
    \email{jose.such@kcl.ac.uk} }}

\definecolor{linkColor}{RGB}{6,125,233}
\hypersetup{%
  pdftitle={\plaintitle},
  pdfauthor={\plainauthor},
  pdfkeywords={\plainkeywords},
  bookmarksnumbered,
  pdfstartview={FitH},
  colorlinks,
  citecolor=black,
  filecolor=black,
  linkcolor=black,
  urlcolor=linkColor,
  breaklinks=true,
}

\begin{document}

\CopyrightYear{2022}
\setcopyright{rightsretained}
\conferenceinfo{CHI'22 Workshop on the Ethics of Conversational User Interfaces,}{April  21, 2022, New Orleans, LA, USA}
\copyrightinfo{Permission to make digital or hard copies of part or all of this work for personal or classroom use is granted without fee provided that copies are not made or distributed for profit or commercial advantage and that copies bear this notice and the full citation on the first page. Copyrights for third-party components of this work must be honored.
For all other uses, contact the owner/author(s). \\ Copyright held by the owner/author(s). \\ CHI’22 Workshop on the Ethics of Conversational User Interfaces, April 21, 2022, New Orleans, LA, USA}

\maketitle
\RaggedRight{} 

\begin{abstract}
Determining how voice assistants should broker consent to share data with third party software has proven to be a complex problem. Devices often require users to switch to companion smartphone apps in order to navigate permissions menus for their otherwise hands-free voice assistant. More in line with smartphone app stores, Alexa now offers “voice-forward consent”, allowing users to grant skills access to personal data mid-conversation using speech. 

While more usable and convenient than opening a companion app, asking for consent ‘on the fly’ can undermine several concepts core to the informed consent process. The intangible nature of voice interfaces further blurs the boundary between parts of an interaction controlled by third-party developers from the underlying platforms. We outline a research agenda towards usable and effective voice-based consent to address the problems with brokering consent verbally, including our own work drawing on the GDPR and work on consent in Ubicomp.
\end{abstract}


\newpage

\section{Introduction}

\begin{marginfigure}[-0pc]
  \begin{minipage}{\marginparwidth}
    \centering
    \includegraphics[width=0.8\marginparwidth]{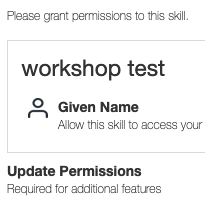}
    \caption{Permissions request card in the Alexa app.}~\label{fig:card}
  \end{minipage}
\end{marginfigure}

\begin{marginfigure}[-0pc]
\begin{minipage}{\marginparwidth}
\section{Sample VFC Flow}
\vspace{1em}
Instead of asking users to enable permissions in the app (and optionally prompting them with a consent card), the skill developer delegates the interaction to the `Alexa Skill'. This is effectively the Alexa operating system, which uses a standardised consent prompt and records the granting or rejection of consent as if the user had used the companion app. \\
\vspace{1em}
\color{orange}
User: Alexa, open Ride Hailer. \\
\color{blue}
Skill: Where are you going? \\
\color{orange}
User: The Space Needle. \\
\color{blue}
Skill: I need access to [...] \\
In order to provide [...] \\
\color{red}
Alexa: Do you give [skill] \\
permission to access [...] \\
You can say 'I approve' or 'no'. \\
\color{orange}
User: I approve. \\
\color{blue}
Skill: The fare will be \pounds10 \\

\end{minipage}
\end{marginfigure}

A vital aspect of the software marketplaces available on today's smart devices, including voice assistants (VAs), is the way that sharing of personal data with third parties is managed. Consent has emerged as the primary means of managing this relationship with skill developers, and is intended to allow users to decide for themselves what they are willing to share when using third party skills\footnote{Even when skills have legitimate interests or contractual justifications for collecting data, consent remains an required \textit{ethical} component of the design and operation of these platforms.}. The growing ubiquity and pervasiveness of these devices---that are often able to listen to everything said in the home---mirrors wider concerns in the Ubicomp community around the information on which users are expected to decide whether or not to grant consent.

Unsurprisingly, designing mechanisms to broker consent for data sharing between users of voice assistants and third-party skills has proven to be somewhat of a wicked problem. When using a skill that requires permission to access personal data, Alexa and Google Assistant direct users to grant access in the companion smartphone app (Figure~\ref{fig:card}). This mirrors the flow of managing permissions on smartphone apps that was once commonplace, where consent would be granted in the app store at the threshold of use. But directing users of a hands-free device that can be used anywhere within earshot to manually interact with a specific device leads to a poor user experience. Last year, Amazon introduced ``voice-forward consent'' (VFC) for Alexa, allowing developers to request consent for data sharing mid-conversation using speech. On the face of it this not only reduces friction in the user experience, but mirrors the shift seen in smartphone apps towards asking for permissions `just in time' (a move recommended by some privacy scholars~\cite{schaub2017designing} but whose implementation has been criticised by others~\cite{shen2021can}).

But the key difference between smartphones and voice assistants---the use of speech as an interaction modality---fundamentally changes the nature of the consent-granting process. In this position paper we outline a research agenda for developing usable and effective voice-based consent for voice assistants, including our own early-stage work on the ethical issues around verbal consent for voice assistants. In this work we draw on literature from HCI and Ubicomp, as well as key data protection regulations including the European General Data Protection Regulation (GDPR)~\cite{euGDPR}, in order to distil out guidelines for ethical verbal consent in voice assistants and other similar technologies. A key motivation for this work is promoting transparency around data collection and use within voice assistant ecosystems; VAs already suffer from a lack of transparency due to the use of speech as an interaction modality, and this is further impacted by a fragmented user experience across assistants, companion apps, and web interfaces.

\section{Can You Meaningfully Consent in 8 Seconds?}
Analysis of the relevant literature and regulation suggests several key issues with the voice forward consent process which we summarise here. These are (1) the introduction of time pressure to the permissions process; (2) difficulty conveying the required amount of information via speech; (3) divorcing the process for granting consent from the process for revoking it; and (4) the lack of distinction between speech from third party skills and the Alexa OS. Each violates established principles of informed consent, albeit in different ways. 

VFC's placement within an interaction adds a previously unseen sense of time pressure to the consent process. Not only does the initiation of VFC halt the present task, therefore after an initial investment of time and effort by the user that will be lost if consent is not granted, but by default Alexa times out and re-prompts the user after eight seconds. While this does not normally end the interaction, it does condition users to provide quick responses to prompts, threatening the freely given nature of informed consent~\cite{10.1145/3411764.3445107}. This is further complicated by the use of specialy crafted consent dialogues on the web that are designed to steer users towards granting consent~\cite{10.1145/3411763.3451230}. Amazon and skill designers have a similar conflict of interest with respect to VFC.

The need to deliver short, concise messages via voice causes further problems. The lower bandwidth of speech compared to graphical user interfaces means that VFC needs to accomplish in a few sentences what would normally occupy paragraphs of text. The sample VFC flow shown on the previous page highlights the brevity employed, mentioning only the name of the skill and permissions requested despite taking around twenty seconds to deliver. Information about the skill developer is omitted, and it is left to the skill to accurately describe (or not) what the information will be used for.

By moving the granting of consent to the voice interface, VFC allows users to grant consent for data sharing without being informed about how they can withdraw consent (or even that they can withdraw consent at all). While clunky, directing users to the companion app which displayed a list of permissions and their status (granted/ not granted) signposted that users could return to the same place in order to revoke consent that they had previously given.

As a final example, though delivery of voice-forward consent is handled by the Alexa operating system there is (by default) no audible difference between speech delivered by the Alexa OS and speech delivered by a third party skill. This makes it difficult for users to accurately interpret and place their trust in VFC dialogues; many will assume that Alexa controls the entire process and oversees the entire consent flow including reasons for processing data. However, it is possible for a skill to imitate the language and response choices used in VFC without delegating to Alexa with no audible difference to the user.

\section{Potential Solutions}
So what can be done? Using the same resources used to highlight the problems with VFC we are exploring a variety of measures and mitigations that can be used by skill and platform developers in their products. At a basic level this involves strengthening policies and mechanisms that are already present, such as incorporating developer justification for using data into the skill certification process and making their disclosure a mandatory part of VFC flows.

For developers, the key first step is to motivate the need for access to data in the precious few words that are available when communicating permission requests to users. While not suggested in the Amazon developer documentation, utilising other media such as response cards and external communications channels can help to overcome the bare-bones nature of VA conversational interactions.

For VA platforms, allowing users to easily distinguish between when their device is speaking on behalf of a skill vs as the operating system is a crucial first step, and facilitates the formation of more accurate mental models by users. Modification of the usual conversational turn-taking model to facilitate engagement with consent and removal of the timeout window could help lessen feelings of being rushed. Pointers to additional resources, perhaps via the companion app, would give users material they can return to after the interaction is complete to foster understanding and reflection. Allowing users to revoke consent via speech, and describing how to do so when it is granted would remove barriers to engagement by restoring the symmetry of consent flows. Where users do revoke consent, platforms could recommend alternative skills that do not require as many permissions.

Finally, the situation also presents opportunities for VAs to embrace better consent practises that will, we believe, also help people feel more confident about using devices that are often perceived as inherently unsettling. Examples of this include the potential to check-in with users in the weeks and months following an initial granting of consent, utilising the lightweight nature of VFC as a way of embracing consent as a living, ongoing practice~\cite{10.1145/2493432.2493446}. The necessary conciseness of speech, while appearing on the surface to be a hindrance actually interacts positively with recommendations to provide ``short, specific privacy notices''~\cite{schaub2017designing}.

\section{Research Agenda}
Much needs to be done in order to develop effective and usable voice-based consent mechanisms for voice assistants and technology with similar conversational interfaces. In our own early-stage work (planned for later in 2022), we are planning a Delphi study that will bring together fellow academics, industry partners, and policy partners to discuss best practices for voice-forward consent. The items in the study will be drawn from key regulation (inc. the GDPR) and existing literature on consent and data sharing permissions in HCI and Ubicomp.

We are currently identifing aspects of regulations and prior work that are potentially relevant to voice forward consent dialogues, both in terms of identifying the core issues, as well as in offering solutions. Following the Delphi methodology~\cite{linstone1975delphi} these will be translated into short, single item statements.

For example, Article 13 of the GDPR ``\textit{Where personal data relating to a data subject are collected from the data subject, the controller shall, at the time when personal data are obtained, provide the data subject with [...] the identity and the contact details of the controller and, where applicable, of the controller’s representative}''~\cite{euGDPR} becomes:

\begin{enumerate}
    \item Voice-based consent should include the identity of the data controller
    \item Voice-based consent should include the contact details of the data controller
\end{enumerate}

These are then conferred upon and rated by experts on axes of relevance to the consent process, as well as anticipated actionability and understandability by users. Over several rounds this process yields stability in opinions (although not necessarily consensus amongst panellists), resulting in a set of guidelines around the use of voice-based consent.

Beyond this, in the longer term research agenda for voice-based consent the usability of such mechanisms is incredibly important. In the aftermath of several well-intentioned but ultimately poorly implemented EU privacy regulations\footnote{Such as for cookies, and more recently around the GDPR.} it is vital that voice-based consent does not become another burden to which people become habituated into dismissing without a second thought. This could involve user studies evaluating different implementations of the guidelines, co-design opportunities with different user groups, and policy recommendations to regulators. We envisage that a combination of these methods will ultimately required to put voice-based consent on a firm footing for the future as VA technology develops and encompasses new aspects of daily life.

\section{Conclusion}
The introduction of voice-forward consent for Alexa represents a great improvement in user experience but one that potentially undermines the informed consent process. We set out a research agenda for developing usable, effective voice-based consent mechanisms. Our own early-stage research involves using literature from HCI and Ubocomp, alongside extracts from the GDPR highlight specific problems with the current VFC implementation and potential solutions. As this work progresses we will draw on expert opinion to further motivate and refine the ideas presented in this position paper, ultimately creating guidelines for developers and platforms that promote healthy and dynamic verbal consent practices in voice assistants and similar conversational interfaces.

\section{Acknowledgements}
This work is funded by the Secure AI Assistants project via Grant EP/T026723/1 from the UK Engineering and Physical Sciences Research Council.

\balance{} 

\bibliographystyle{SIGCHI-Reference-Format}
\bibliography{main}

\end{document}